\documentclass[aps,prb,twocolumn,superscriptaddress]{revtex4-1}

\usepackage{amsmath}
\usepackage{empheq}
\usepackage{graphicx}
\usepackage{natbib}

\newcommand{\vv}[1]{\vec{#1}}

\newcommand{\vvm}{\vv{m}}
\newcommand{\vvM}{\vv{M}}

\newcommand{\vvH}{\vv{H}}
\newcommand{\vvHa}{\vv{H}_{\mathrm{app}}}

\newcommand{\Msat}{M_{\mathrm{s}}}

\newcommand{\vvHe}{\vv{H}_{\mathrm{exch}}}
\newcommand{\vvHd}{\vv{H}_{\mathrm{demag}}}
\newcommand{\vvHpin}{\vv{H}_{\mathrm{pin}}}
\newcommand{\Upin}{U_{\mathrm{pin}}}

\newcommand{\japp}{j_{\mathrm{app}}}

\newcommand{\DE}{\Delta E}
\newcommand{\DEmax}{\Delta E_{\mathrm{max}}}

\newcommand{\mean}[1]{\langle #1 \rangle}
\newcommand{\avg}[1]{\langle #1 \rangle}
\newcommand{\dg}{^{\circ}}
\newcommand{\kbt}{k_{\mathrm{B}} T}

\newcommand{\Hcrit}{H_{\mathrm{crit}}}
\newcommand{\jcrit}{j_{\mathrm{crit}}}
\newcommand{\Icrit}{I_{\mathrm{crit}}}
\newcommand{\pinang}{\theta_{\mathrm{P}}}
\newcommand{\dwpos}{x_{\mathrm{DW}}}
\newcommand{\dwvel}{v_{\mathrm{DW}}}

\newcommand{\derivp}[2]{\partial_{#2} #1}

\begin{document}

\title{Enhanced spin transfer torque effect for transverse domain walls
       in cylindrical nanowires}

\author{Matteo~Franchin}
\email[]{franchin@soton.ac.uk}
\affiliation{School of Engineering Sciences, University of Southampton,
             SO17 1BJ Southampton, United Kingdom}

\author{Andreas~Knittel}
\affiliation{School of Engineering Sciences, University of Southampton,
             SO17 1BJ Southampton, United Kingdom}

\author{Maximilian Albert}
\affiliation{School of Engineering Sciences, University of Southampton,
             SO17 1BJ Southampton, United Kingdom}

\author{Dmitri Chernyshenko}
\affiliation{School of Engineering Sciences, University of Southampton,
             SO17 1BJ Southampton, United Kingdom}

\author{Thomas~Fischbacher}
\affiliation{School of Engineering Sciences, University of Southampton,
             SO17 1BJ Southampton, United Kingdom}

\author{Anil~Prabhakar}
\affiliation{Department of Electrical Engineering, IIT Madras,
             Chennai 600036, India}

\author{Hans~Fangohr}
\affiliation{School of Engineering Sciences, University of Southampton,
             SO17 1BJ Southampton, United Kingdom}

\date{\today}

\begin{abstract}
Recent studies have predicted extraordinary properties for transverse
domain walls in cylindrical nanowires: zero depinning current, the
absence of the Walker breakdown, and applications as domain wall
oscillators. In order to reliably control the domain wall motion, it
is important to understand how they interact with energy barriers,
which may be engineered for example through modulations in the
nanowire geometry (such as notches or extrusions) or as
inhomogeneities in the material's crystal anisotropy.

In this paper, we study the motion and depinning of transverse domain
walls through potential barriers in ferromagnetic cylindrical
nanowires. We use (i) magnetic fields and (ii) spin-polarized currents
to drive the domain walls along the wire. Barriers are modelled as a
section of the nanowire which exhibits a uniaxial crystal anisotropy
where the anisotropy easy axis and the wire axis enclose a variable
angle $\pinang$.
Using (i) magnetic fields, we find that the minimum
and the maximum fields required to push the domain wall through
the barrier differ by $30\%$.
On the contrary, using (ii) spin-polarized currents,
we find variations of a factor 130 between the minimum value
of the depinning current density (observed for $\pinang=0\dg$,
i.e. anisotropy axis pointing parallel to the wire axis)
and the maximum value (for $\pinang=90\dg$,
i.e. anisotropy axis perpendicular to the wire axis).

We study the depinning current density
as a function of the height of the energy barrier using numerical
and analytical methods.
We find that for an industry standard energy barrier of $40\,\kbt$,
a depinning current density of about $5\,\mu A$
(corresponding to a current density of $6\times10^{10}\,\mathrm{A/m}^2$
in a nanowire of $10\,\mathrm{nm}$ diameter)
is sufficient to depin the domain wall.

We reveal and explain the mechanism that leads to these unusually low
depinning currents. One requirement for this new depinning mechanism
is for the domain wall to be able to rotate around its own axis.
With the right barrier design,
the spin torque transfer term is acting exactly against the damping in
the micromagnetic system, and thus the low current density is
sufficient to accumulate enough energy quickly.
These key insights may be crucial
in furthering the development of novel memory technologies, such as the
racetrack memory, that can be controlled through low current densities.

\end{abstract}

\maketitle

\section{Introduction}
The current induced motion of domain walls (DWs) in ferromagnetic nanowires
has been the subject of intense study in recent years. Being able to move and
control DWs accurately is an important step towards the realisation of
devices such as the racetrack memory \cite{Parkin2008}. Two main challenges are
the high critical current density, $j_{\mathrm{D}}$, required to depin the DW
and initiate the DW motion \cite{Tatara2004} and the so-called Walker breakdown,
a phenomenon which limits the DW velocity and is caused by deformations
in the magnetization structure \cite{Klaui2005}.
It has been recently observed \cite{Yan2010, Wieser2010} that transverse
domain walls (TDWs) in cylindrical nanowires are not subject to such
limitations: $j_{\mathrm{D}}$ is zero, and the Walker breakdown does not
occur. These DWs are able to propagate uniformly, without any deformation
of their internal structure.
Moreover, recent theoretical studies\cite{franchin2008jap,Franchin2008prb,Ono2008}
suggest that TDWs may play an important role in the emerging research area
of domain wall oscillators\cite{He2007,Matsushita2009,Bisig2009,
Finocchio2010,Masaaki2011} (DWOs), providing an effective way to sustain
magnetization dynamics with direct currents.

The main obstacle for the experimental observation
of these effects is that TDWs only occur in cylindrical nanowires with small
diameter (below $\sim50\,\mathrm{nm}$ for Permalloy). In wires with greater diameter
the demagnetizing field is strong enough to force the magnetization to follow
the cylindrical surface of the wire, thus changing the internal structure of
the DW, giving rise to a vortex DW\cite{Wieser2004,Yan2010}.
As a consequence, the experimental validation of the predicted properties
of TDW requires the fabrication of cylindrical nanowires with diameter below
$50\,\mathrm{nm}$, which is challenging from an experimental point of view.
Recently, however, novel techniques have been developed for fabricating
cylindrical nanowires with small diameter
\cite{Cao2001,Ruitao2007,Pitzschel2011a}.
Ruitao \emph{et. al.}\cite{Ruitao2007} describe a procedure to obtain
single crystalline Permalloy nanowires with an average diameter
of $\sim 30\,\mathrm{nm}$, through in situ filling of the inner cavities
of carbon nanotubes.
Pitzschel \emph{et. al.}\cite{Pitzschel2011a} discuss the fabrication
of arrays of Ni cylindrical nanowires with diameter between $80$ and $160\,\mathrm{nm}$,
using electrodeposition.
Such arrays are technologically relevant, as they may represent a practical
and convenient way to densely pack nanowires for memory storage applications.
In particular, the authors discuss how the diameter of the nanowires
can be modulated in order to introduce geometrical ``artifacts''
which act as barriers for the propagation of DWs.
These barriers may be necessary in order to control
the DW motion and make it more predictable and reliable.
It is then important for technological applications to study and understand
the motion of TDWs through obstacles (which may be realized
as inhomogeneities in the material or in the geometry in the nanowire).

In this paper we present a micromagnetic study of the field-driven
and current-driven motion of transverse domain walls through
potential barriers in cylindrical nanowires.
We consider two kinds of barriers: cylindrically symmetric barriers,
whose associated energy term is invariant for rotations of the magnetization
around the nanowire axis, and cylindrically asymmetric barriers,
whose energy term does depend on the rotational state of the magnetization.
When passing through a cylindrically symmetric barrier the TDW is
``rotationally-free'', i.e. it can rotate freely around the nanowire axis.
On the contrary, when passing through a cylindrically asymmetric barrier
the TDW is ``rotationally-bound'', as the barrier introduces an energy penalty
for rotations of the DW.
We find that the current density required to push the DW through the barrier
is radically different in the two cases.
In particular, we report that current densities
of $10^{10}-10^{11}\, \mathrm{A/m}^2$ can be used to depin
rotationally-free DWs from barriers which are thermally ``stable'' at room temperature
(the associated energy is greater than $40\,\kbt$ at $T=300\,\mathrm{K}$).
In contrast, depinning a rotationally-bound DW requires current densities
which are higher by a factor $\sim130$.

\section{The system}
\begin{figure}
\includegraphics[width=8.5cm]{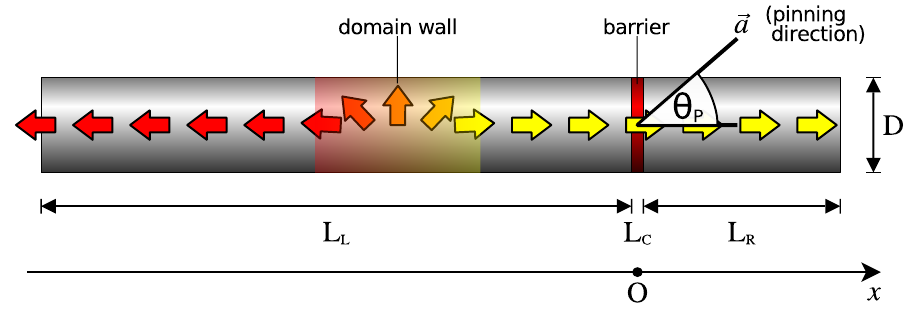}
\caption{\label{fig_system} (Color online) Sketch of the system.
         The arrows on the cylinder axis represent
         the magnetization. The transverse DW on the left is pushed
         through the barrier on the right.
         The barrier is modeled as a region where an additional uniaxial
         anisotropy pins the magnetization along the direction determined
         by $\pinang$.}
\end{figure}

In order to study the field-driven and current-driven motion
of a DW through a potential barrier, we employ the system
schematically shown in Fig. \ref{fig_system}, a long cylindrical ferromagnetic
nanowire.
The potential barrier is modeled as a thin region of the wire
of length $L_{\mathrm{C}}$ where a uniaxial anisotropy pins the magnetization
along a given direction $\vv{a}$ with a given strength $K_1$.
In this first part of the paper, we fix the value of the anisotropy
constant $K_1$ and study how the motion of the DW depends on $\pinang$,
the angle between the unit vector $\vv{a}$ and the nanowire axis.

The nanowire is subdivided into three regions:
one longer region on the left with length $L_{\mathrm{L}}$
and one shorter region on the right with length
$L_{\mathrm{R}}$, separated by the barrier,
a thin layer of width $L_{\mathrm{C}}$.
In the simulation we ``place'' a DW on the left of the barrier
and try to push it to the right using increasing
applied fields (Sec. \ref{sec:appfield})
or current densities (Sec. \ref{sec:appcurrent}).

The dynamics of the magnetization, $\vvM$,
is computed using the Landau-Lifshitz equation, extended with two
additional terms in order to model spin transfer torque
effects \cite{ZhangLiModel2004}:
\begin{eqnarray}
\derivp{\vvM}{t} & = &
    -\gamma \, \vvM \times \vvH
    +\frac{\alpha}{\Msat} \, \vvM \times \derivp{\vvM}{t} \nonumber\\
& & {} + v \, \derivp{\vvM}{x}
    -\frac{\xi v}{\Msat} \, \vvM \times \derivp{\vvM}{x}
\label{eq:llgzl}
\end{eqnarray}
In this equation, $\Msat = \|\vvM\|$ is the saturation magnetization,
$\vvH$ is the effective magnetic field,
$\gamma$ is the gyromagnetic ratio, $\alpha$ is the damping parameter.
The current density, $\japp$, is applied along the $x$ direction
and enters the model through the parameter
$v=\frac{P \japp \mu_B}{e \Msat (1+\xi^2)}$, where $P$
is the degree of polarization of the spin current,
$\mu_B$ is the Bohr magneton, $e$ the absolute value
of the electron charge, $\xi$ the non-adiabatic parameter.
Note that in this paper we reason in terms of fully polarized current
densities, $j = P \japp$, rather than in terms of the actual applied
current densities, $\japp$.

The effective field is calculated as
$\vvH = \vvHe + \vvHd + \vvHa + \vvHpin$,
where $\vvHe = \frac{2A}{\mu_0 \Msat}\,\nabla^2 \vvm$
is the exchange field, $\vvHd$ is the magnetostatic field,
$\vvHa$ is the applied field
and $\vvHpin = (2 K_1 \vv{m} \cdot \vv{a} /\mu_0 \Msat) \, \vv{a}$
is the magnetic anisotropy corresponding to the pinning potential
$\Upin = - K_1 \, (\vv{m} \cdot \vv{a})^2$.
$K_1$ is the only parameter which varies in space:
it is zero outside the barrier region, and is set to
$K_1 = 0.5\times10^6\,\mathrm{J}/\mathrm{m}^3$ inside it.
The other parameters are homogeneous throughout the whole nanowire
and are set to $\Msat = 0.86 \times 10^6 \, \mathrm{A/m}$,
$A = 1.3 \times 10^{-11} \, \mathrm{A/m}$,
$\gamma = 2.21 \times 10^5 \, \mathrm{m}/(\mathrm{A s})$,
$\alpha = 0.01$ and $\xi = 0.01$, which are typical values for Permalloy.
Concerning the geometry of the cylindrical nanowire we take
$L_{\mathrm{L}}= 148.5\,\mathrm{nm}$,
$L_{\mathrm{R}} = 48.5\,\mathrm{nm}$,
$L_{\mathrm{C}} = 3\,\mathrm{nm}$,
so that the total length is
$L = L_{\mathrm{L}} + L_{\mathrm{C}} + L_{\mathrm{R}} = 200\,\mathrm{nm}$.
The diameter is chosen as $D = 10\,\mathrm{nm}$.
Contributions from the Oersted field and the Joule heating are
negligible for nanowires of small radius\cite{Franchin2008prb}
and are thus not included in the model.

We note that, when $\pinang > 0\dg$, the pinning field $\vvHpin$
tries to pull the magnetization out of the axis of the nanowire.
The magnetostatic field, however, opposes this and
--- for the value of $K_1$ considered in this paper ---
manages to keep the magnetization along the axis of the nanowire,
even inside the potential barrier.

Eq. \eqref{eq:llgzl} is discretized over a finite element mesh
and the simulations are carried out using a version
of the micromagnetic simulation package
\emph{Nmag} \cite{Fischbacher2007a,nmag2006},
extended to model spatially varying magnetic anisotropies.

\section{Field-driven DW motion} \label{sec:appfield}
We carry out a number of simulations to determine the critical
field, $\Hcrit$, which must be applied (in the direction of the nanowire
axis) in order to push the DW through the potential barrier.
In particular, we fix the pinning strength to
$K_1 = 0.5 \times 10^6\,\mathrm{J}/\mathrm{m}^3$
and perform one simulation for each value of the pinning angle $\pinang$,
which is changed from zero (pinning along the nanowire axis) to $90\dg$
(pinning orthogonal to the nanowire axis) in steps of $5\,\dg$.

\begin{figure}
\includegraphics[width=8.7cm]{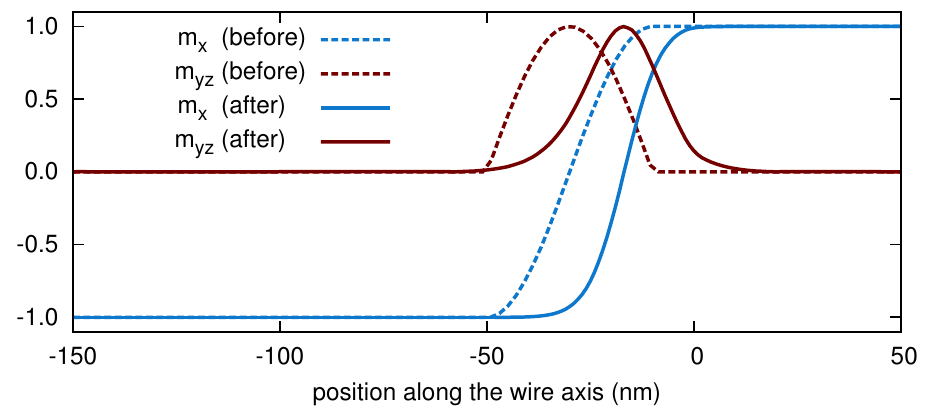}
\caption{\label{fig_dwprof} (Color online)
  The components of the normalized magnetization,
  $\vvm = \vvM/\Msat$, along the axis of the nanowire
  before (dashed lines) and after (solid lines) the preliminary relaxation
  computation.
  The blue curves (light in greyscale) show the $x$ (axial) component
  of $\vvm$, while the dark red curves (dark in greyscale) show the
  component of $\vvm$ orthogonal to the axis.
}
\end{figure}

Each simulation is carried out in two parts. In the first part,
a preliminary relaxation computation is carried out and the system
settles into a metastable state where a DW is located on the left of the barrier.
In the second part, the applied field is increased gradually
until the DW passes through the barrier.
For the preliminary relaxation we set the initial magnetization
in the following way,
\begin{eqnarray}
\vvM(x, y, z) & = & \Msat \, (\sin \theta(x),\, \cos \theta(x),\, 0),
\end{eqnarray}
where $\theta(x) = \frac{\pi}{2} \min \left(1.0,\,
\max \left(-1.0,\, \tiny (x - x_0)/w\right) \right)$
and $x_0 = -30\,\mathrm{nm}$, $w = 20\,\mathrm{nm}$
(the reference system is chosen as in Fig. \ref{fig_system}).
For this choice of parameters,
the magnetization relaxes into a tail-to-tail DW,
located on the left of the nanowire,
similarly to what is schematically shown in Fig. \ref{fig_system}.
The components of the magnetization along the axis of the nanowire before
and after the relaxation computation are shown in Fig. \ref{fig_dwprof}.
The relaxation is carried out using an artificially high value
for the damping, $\alpha = 0.5$, to speed up the computation,
and is done in the presence of a small magnetic field,
$H_0 = 10^3\,\mathrm{A}/\mathrm{m}$,
which is applied along the negative $x$ direction
and is used to push the DW towards the barrier.
We want indeed to ensure that the DW settles into an equilibrium
position against the barrier (e.g the center of the barrier) before
starting to ``depin'' it.
The convergence criterion used to stop the simulation is the following:
\begin{equation}
\max \left\{ \frac{1}{\Msat} \left\| \frac{\mathrm{d} \vvM_i}{\mathrm{d} t} \right\|
     \right\}_i < \varepsilon
\label{eq:convergence}
\end{equation}
where the index $i$ runs over all the sites of the finite element mesh
and $\varepsilon = 1\,\mathrm{deg}/\mathrm{ns}$.

\begin{figure}
\includegraphics[width=8.0cm]{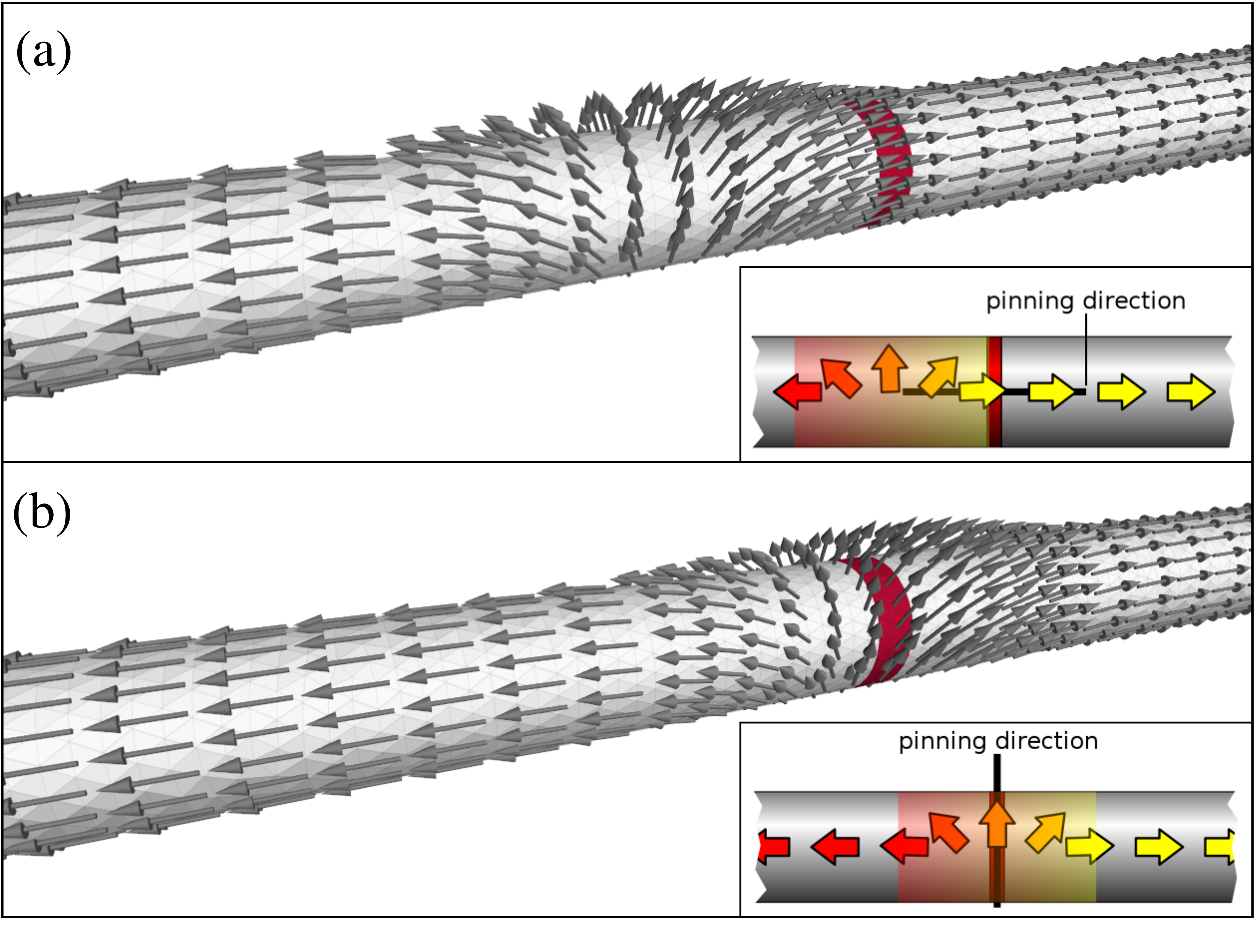}
\caption{\label{fig_m0s} (Color online)
  The relaxed magnetization configurations for $\pinang = 0\dg$ (a)
  and $\pinang = 90\dg$ (b). The sketches on the corners of the two
  figures show that the DW moves into the barrier to allow the magnetization
  to align along the pinning direction in the barrier.
}
\end{figure}

The procedure described above brings the system to a state
which is different for different values of the pinning angle, $\pinang$.
In particular, when $\pinang$ is zero the DW approaches
the barrier from the left without penetrating it (Fig. \ref{fig_m0s}(a).),
as this would require the magnetization in the region to move out
of the pinning direction.
On the other hand, when $\pinang$ is $90\dg$ the DW
``falls'' into the center of the barrier allowing the magnetization
to align along the pinning direction, with a consequent reduction in energy
(see Fig. \ref{fig_m0s}(b)).
Depending on $\pinang$ we then have two different scenarios,
where the barrier either repels the DW (as for $\pinang = 0\dg$)
or attracts it (as for $\pinang = 90\dg$).

The magnetization configuration obtained from the preliminary relaxation
computation is used as the initial state for the second part
of the simulation, in which the damping is set to the realistic value,
$\alpha = 0.01$, and the applied field is gradually increased
until the DW passes completely through the potential barrier.
In particular, we increase the field
in steps of $\Delta H = 10^2\,\mathrm{A}/\mathrm{m}$ and for each value
of the applied field we relax the system, i.e. perform a time integration
of Eq. \eqref{eq:llgzl} until the convergence criterion \eqref{eq:convergence}
is met.
To determine when the DW passes through the barrier and stop the simulation,
we analyze the values of the spatially averaged
normalized magnetization $\mean{\vvm}$, where $\vvm = \vvM/\Msat$.
The $x$ component of this vector, $\mean{m_x}$, provides a good indication
of the position of the DW, while the $y$ and $z$
components provide information about its state of rotation.
In particular, if we assume the DW to be mirror-symmetric along the $x$ axis,
we get $\mean{m_x} \approx (L_{\mathrm{R}} - L_{\mathrm{L}} - 2\dwpos)/L$,
where $\dwpos$ is the position of the DW, while $L$ is the total length
of the nanowire, $L = L_{\mathrm{L}} + L_{\mathrm{C}} + L_{\mathrm{R}}$.
In the simulations we assume the DW to pass through the potential barrier
when $\mean{m_x} < -0.75$, which corresponds, following the formula above,
to a position of $\dwpos = 25\,\mathrm{nm}$.

\begin{figure}
\includegraphics[width=8.0cm]{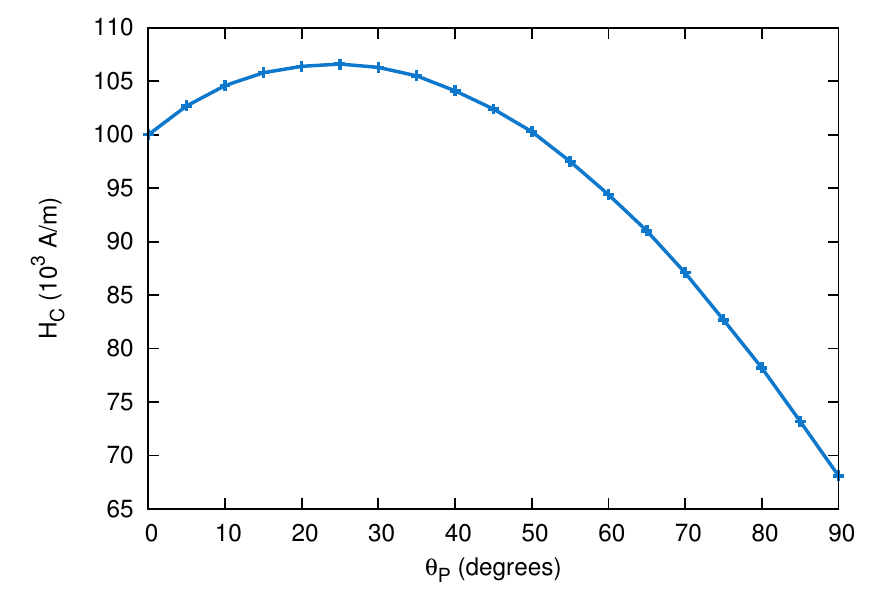}
\caption{\label{fig_hcrit} (Color online)
  $H_{\mathrm{crit}}$, the field required in order to push the DW
  throughout the potential barrier, is plotted as a function of the pinning
  angle, $\pinang$.}
\end{figure}

The results of the simulations are collected in Fig. \ref{fig_hcrit}.
\begin{figure}
\includegraphics[width=8.0cm]{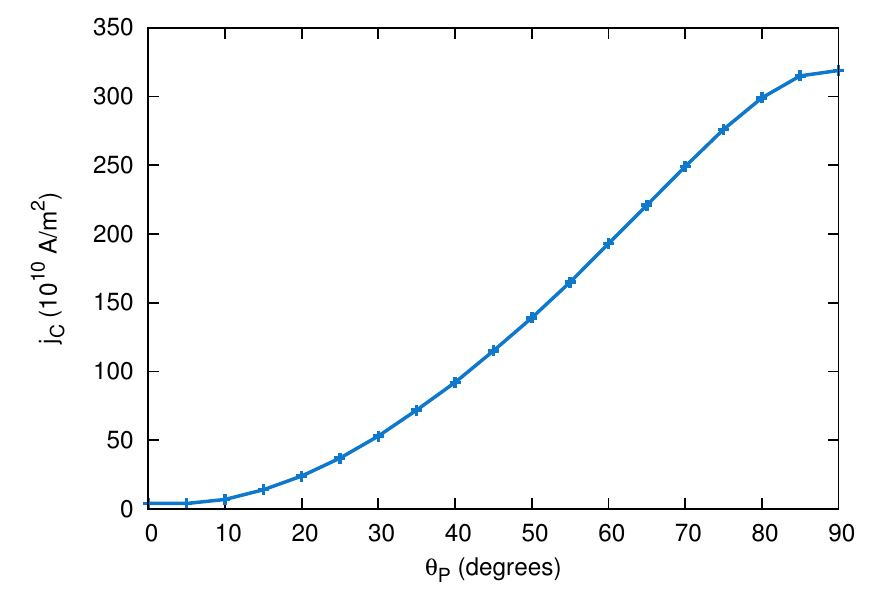}
\caption{\label{fig_jcrit} (Color online)
  $j_{\mathrm{crit}}$, the current required in order to push the DW
  through the potential barrier, is plotted as a function of the pinning
  angle, $\pinang$.}
\end{figure}
The field necessary to move the DW through the barrier varies
between $(68.05 \pm 0.05)$ and $(106.55 \pm 0.05)\,\mathrm{kA}/\mathrm{m}$.

\section{Current-driven DW motion} \label{sec:appcurrent}
In order to determine the current necessary to push the DW
through the barrier we follow a procedure analogous to the one described
in the previous section, with two main differences.
First, to push the DW we use a current rather than a field.
Second, we abandon the convergence criterion in Eq. \eqref{eq:convergence}.
This convergence criterion works well in the field-driven case
because the torque is guaranteed to decrease in time due to the damping term.
In the current-driven case, however, the spin transfer torque can oppose
the damping and bring the system to a steady state where the magnetization
dynamics is sustained in time.
In particular, it may happen that the DW reaches a state where it does
not advance anymore along the wire but keeps rotating around its
axis \cite{Franchin2008prb}.
In this situation, the torque never falls below the given threshold,
$\varepsilon$, and Eq. \eqref{eq:convergence} is never satisfied.
We then use a convergence criterion which stops the simulation when
the DW translational velocity (along $x$) falls below a certain threshold.
For convenience, we formulate it in terms of $\mean{m_x}$, as follows:
\begin{equation}
\left| \frac{\mean{m_x(t + \Delta t)} - \mean{m_x(t)}}{\Delta t} \right|
  < \varepsilon,
\label{eq:jconvergence}
\end{equation}
with $\varepsilon = 10^{-5} \, \mathrm{ns}^{-1}$. The convergence
criterion is checked every $\Delta t = 100\,\mathrm{ps}$ and the simulation
is stopped when it is satisfied for five consecutive times.
We note that Eq. \eqref{eq:jconvergence} may be expressed in terms
of the DW velocity using the relation
$\mean{m_x} = (L_{\mathrm{R}} - L_{\mathrm{L}} - 2\dwpos)/L$,
\begin{equation}
\dwvel =
\left| \frac{\dwpos(t + \Delta t) - \dwpos(\Delta t)}{\Delta t} \right|
  < \frac{L}{2} \, \varepsilon,
\end{equation}
which reduces, in our specific case, to
$\dwvel < 10^{-3}\,\mathrm{nm}/\mathrm{ns}$.
We also pin the magnetisation at the left and right ends of the simulated wire
to reduce finite-size effects (such as the nucleation of DW, which may
occur at current densities above $\sim 10^{12}\,\mathrm{A/m}^2$).

As in the field-driven case, we carry out one simulation
for each chosen value of $\pinang$ and each simulation is subdivided
into a preliminary part, where the DW is prepared in its initial
state on the left of the barrier, and a main part, where the current density
is increased until the DW passes through the barrier.
In the preliminary simulation we use a current density
$j_0 = 10^{10}\,\mathrm{A}/\mathrm{m}^2$ to push the DW from the left
to the right against the potential barrier. The current is directed
along the negative $x$ direction.
We use a high damping, $\alpha = 0.5$, similarly to the field-driven case.
In the main part of the simulation, we restore the damping to the more
realistic value $\alpha = 0.01$ and increase the current density in steps
of $\Delta j = 10^{10}\,\mathrm{A}/\mathrm{m}^2$.
The results of the simulations are shown in Fig. \ref{fig_jcrit}.

We see that the current density required to push the DW through the potential
barrier, $\jcrit$, increases rapidly as the pinning direction moves out
of the axis of the cylinder: for $\pinang = 0\dg$ we get
$\jcrit = (0.0239 \pm 0.0001) \times 10^{12}\,\mathrm{A}/\mathrm{m}^2$
(this value was determined with a separate simulation to get improved
accuracy), while for $\pinang = 90\dg$ we get
$\jcrit = (3.185 \pm 0.005) \times 10^{12}\,\mathrm{A}/\mathrm{m}^2$.
We conclude that the critical current $\jcrit$ varies roughly by a factor
$130$ depending on the direction along which the barrier
pins the magnetization. This is a quite remarkable feature
considering what happens in the field-driven case, where $\Hcrit$ does
actually decrease for pinning angles approaching $90\dg$.

To understand why the value of $\pinang$ has such a strong influence
on the critical current $\jcrit$ we have to analyze the dynamics
of the magnetization.
\begin{figure}
\includegraphics[width=8.5cm]{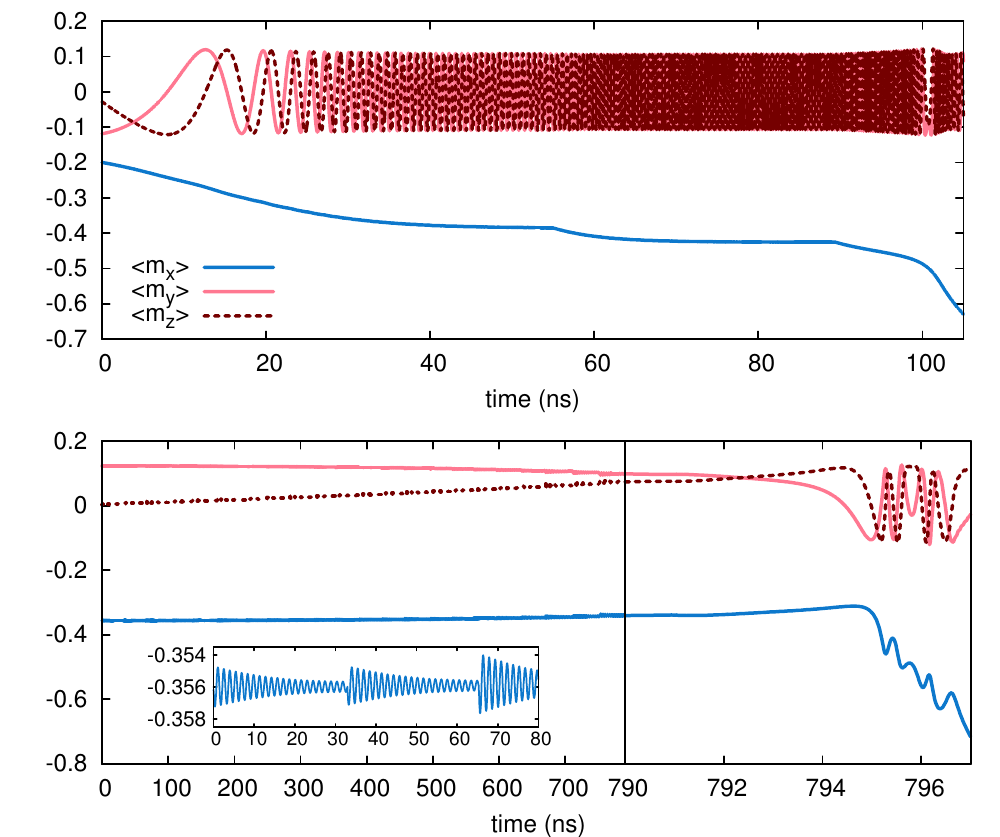}
\caption{\label{fig_jdyn} (Color online) Comparison of the current driven
  dynamics of the spatially averaged magnetization, $\mean{\vvm}$,
  for $\pinang = 0\dg$ (top) and $\pinang = 20\dg$ (bottom).
  The inset in the bottom plot shows the oscillatory motion of the DW induced
  by the pinning in the barrier. The discontinuities in $\mean{m_x}$
  correspond to changes in the applied current density
  (see Fig. \ref{fig_appj}) following the procedure described in the text.
  Two different scales are used for the time axis of the bottom plot
  in order to give clear information on the dynamics during the two different
  phases of the switching process.
  Note that the dynamics of the preliminary relaxations is omitted in
  these figures.}
\end{figure}
\begin{figure}
\includegraphics[width=8.5cm]{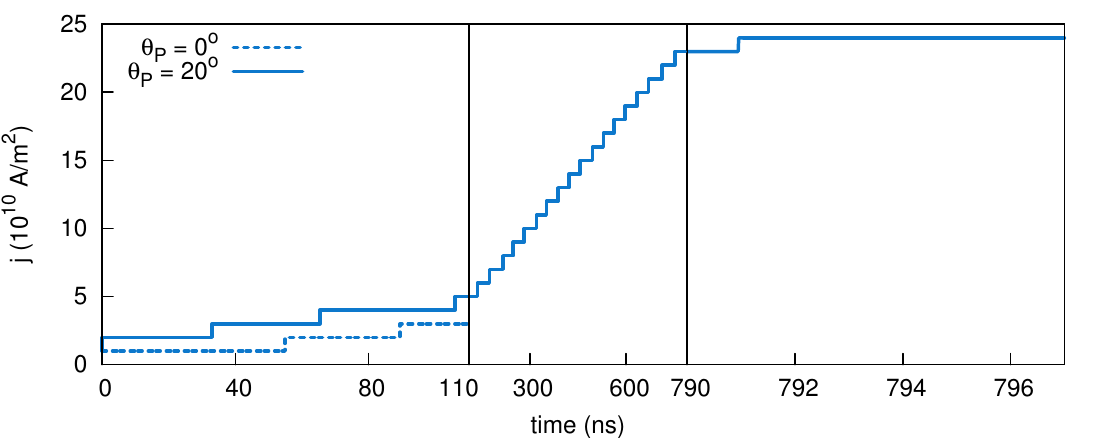}
\caption{\label{fig_appj} (Color online) The applied current density
  for the two simulations of Fig. \ref{fig_jdyn} and Fig. \ref{fig_endyn}.
  The vertical lines in the plot mark a change in the scale of the time axis.}
\end{figure}
Fig. \ref{fig_jdyn} shows the time evolution of the components
of the spatially averaged magnetization, $\mean{\vvm}$,
during the simulations for $\pinang = 0\dg$ (top)
and $\pinang = 20\dg$ (bottom).
When $\pinang = 0\dg$, the DW rotates around the nanowire axis,
as shown by the $y$ and $z$ components of $\mean{\vvm}$.
At the same time, the $x$ component shows that the DW gradually compresses
against the barrier as the current density, $j$, is increased
from the value $10^{10}\,\mathrm{A/m^2}$,
to $2\times10^{10}\,\mathrm{A/m^2}$ and $3\times10^{10}\,\mathrm{A/m^2}$
at times $0\,\mathrm{ns}$, $54.9\,\mathrm{ns}$ and $89.4\,\mathrm{ns}$,
respectively (dashed line in Fig. \ref{fig_appj}).
The effect of the spin transfer torque is constantly accumulating
in time and is stored in form of barrier penetration and DW compression.

In the case $\pinang = 20\dg$, the potential barrier pins the
magnetization out of the direction of the nanowire axis,
thus breaking the rotational symmetry of the system (i.e. the energy
of the system is not invariant for rotations of the magnetization).
As a consequence, the DW cannot freely rotate around the axis
of the nanowire, but rather precesses narrowly around its equilibrium
configuration.
The inset at the bottom of Fig. \ref{fig_jdyn} confirms
that $\mean{m_x}$ oscillates weakly and does not vary
significantly during the application of increasing current densities
(solid line in Fig. \ref{fig_appj}).
Only at time $791.0\,\mathrm{ns}$, when the current density reaches
the critical value, $\jcrit = 24\times10^{10}\,\mathrm{A/m^2}$,
the DW starts to penetrate significantly into the barrier.
In then takes about $2\,\mathrm{ns}$ for the DW to pass completely through it.

\begin{figure}
\includegraphics[width=7.8cm]{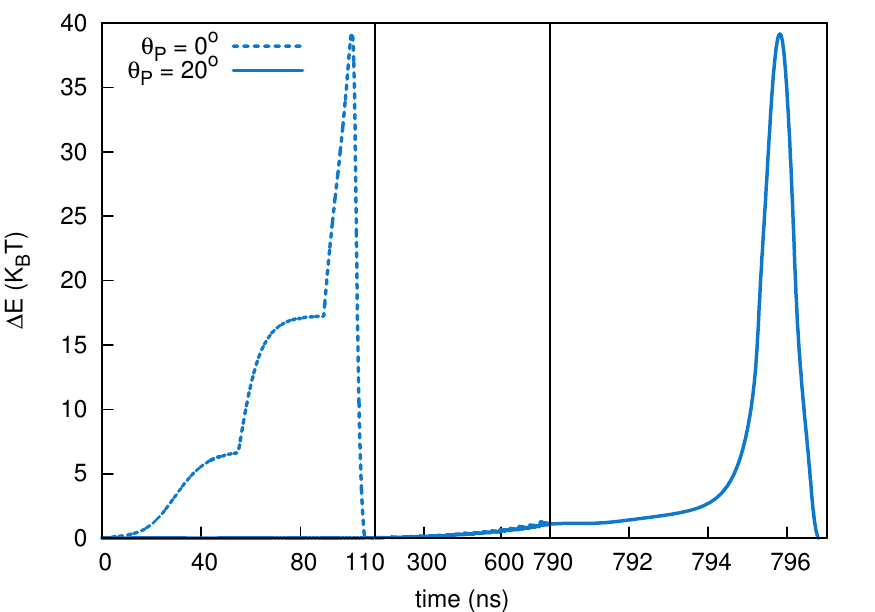}
\caption{\label{fig_endyn} (Color online)
  Evolution of the energy for the two cases $\pinang=0\dg$ (dashed line)
  and $\pinang=20\dg$ (solid line) considered in Fig. \ref{fig_jdyn}
  and \ref{fig_appj}.
  Units of $\kbt = 4.14 \times 10^{-21}\,\mathrm{J}$ ($T = 300\,K$) are used
  for the energy. The time axis uses different scales as in
  Fig. \ref{fig_appj}.}
\end{figure}
Fig. \ref{fig_endyn} shows the micromagnetic energy, $\DE$, in the two
cases $\pinang = 0 \dg$ (dashed line) and $\pinang = 20 \dg$ (solid line).
$\DE(t)$ is computed as $E(t) - E(0)$, where $E(t)$ is the energy of the system
at time $t$.
The height of the energy barrier, $\DEmax$, is almost the same in the
two cases.
In particular, for $\pinang = 0\dg$, $\DEmax = 39.24\,\kbt$
while for $\pinang = 20\dg$, $\DEmax = 39.16\,\kbt$.
It is interesting to notice that in the case of a rotationally symmetric
barrier, the energy increases gradually as a response to the increase
of current density.
On the contrary, in the case of an asymmetric barrier, the energy
changes very weakly with $j$.
In particular, the energy changes less than $1.2\,\kbt$
when $j$ is increased from $10^{10}\,\mathrm{A/m^2}$
to $23\times10^{10}\,\mathrm{A/m^2}$.
At time $t = 791\,\mathrm{ns}$, however, the current is increased to the value
$\jcrit = 24\times10^{10}\,\mathrm{A/m^2}$ and it takes only $5\,\mathrm{ns}$
for the DW to pass through the remaining part of the barrier.

\section{Discussion}
To understand why the rotationally symmetric barrier can be penetrated
much more easily than the asymmetric one, we start from Eq. \eqref{eq:llgzl}.
We express the equation in the form where the time derivative
of $\vvm$ appears only on the left hand side (this requires a substitution
of the equation in itself) and both of its sides have been divided by $\gamma \Msat$:
\begin{eqnarray}
\frac{1}{\gamma'} \derivp{\vvm}{t} & = &
    -\vvm \times \vvH
    -\alpha \, \vvm \times (\vvm \times \vvH) \nonumber\\
& & {} + (1 + \alpha \xi) \, \frac{v}{\gamma} \, \derivp{\vvm}{x} \nonumber\\
& & {} - (\xi - \alpha) \, \frac{v}{\gamma} \, \vvm \times \derivp{\vvm}{x},
\label{eq:llgzl_explain}
\end{eqnarray}
where $\gamma' = \gamma/(1 + \alpha^2)$ and $\vvm = \vvM/\Msat$
is a unit vector.
The right hand side of this equation receives the contributions of four terms.
They are, respectively, the precession term, the damping term,
the adiabatic Spin Transfer Torque (STT) and the non-adiabatic STT.
The strongest of these terms is typically the precession term,
which receives contributions of the order of $10^6\,\mathrm{A/m}$
(from the applied field, the demagnetizing field, the magneto-crystalline
anisotropy) or higher (from the exchange field).
The damping term has lower magnitude, due to the prefactor $\alpha$.
For Permalloy, $\alpha \approx 0.01$ and the damping term is then
100 times weaker than the precession term.
Despite this, the damping term has a very important role for the physics
of the system and has substantial impact on its dynamics.
The damping is very effective because it keeps changing the magnetization
towards the direction that yields the highest reduction
in energy.\footnote{By definition, $\vv{H}$ is pointing opposite
to the functional derivative of the micromagnetic energy, $E$, and hence
towards the direction which minimizes $E$.
From Eq. \eqref{eq:llgzl_explain}, the damping term is
$-\alpha \, \vvm \times (\vvm \times \vvH)
= \alpha \vvH - \alpha (\vvm \cdot \vvH) \, \vvm$, which is $\alpha \vvH$
minus the component which would change the magnitude of the magnetization.
In other words, the damping term points towards the direction
which yields the largest reduction in energy, subject to the constraint
$\vvm^2 = 1$.}
Its effect, hence, accumulates in time and brings the system to a local
energy minimum.

On the other hand, the first STT term (the second STT term is smaller
of a factor $\xi - \alpha$ and will be neglected in this analysis)
does not have this property: while it can have magnitude similar to
the damping term (for Permalloy and $j = 10^{12}\,\mathrm{A/m^2}$, the
adiabatic term can bring contributions of the order of $10^4\,\mathrm{A/m}$
in Eq. \eqref{eq:llgzl_explain}), it does not always point in the direction
which is best suited to pump energy into the system.
Often, this term acts in the direction opposite to the much stronger
precession term, with the result that the STT pumps energy in and out
in an alternate fashion, i.e. there is no ``accumulation'' of the effect.

\begin{figure}
\includegraphics[width=8.0cm]{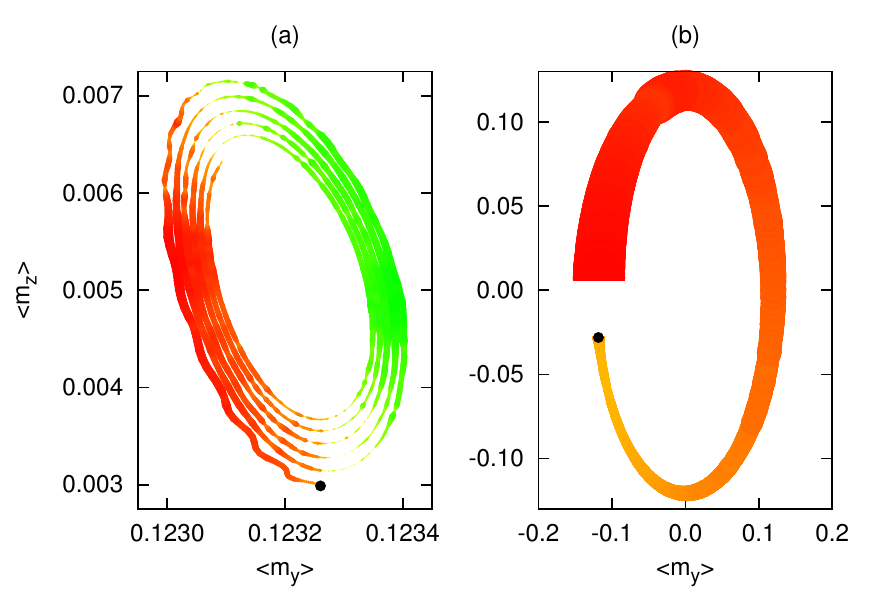}
\caption{\label{fig_enpump} (Color online)
  Current driven trajectories of the average magnetization, $\avg{\vv{m}}$,
  projected on the $yz$-plane (orthogonal to the nanowire axis).
  The dynamics is in response to the application of a current density
  $j=2 \times 10^{10}\,\mathrm{A/m^2}$ in the case of asymmetric barrier,
  $\pinang = 20\dg$ (a) and symmetric barrier, $\pinang = 0\dg$ (b).
  The line thickness is proportional to the magnitude of the energy which
  is being pumped in (red in color, dark in greyscale) or out
  (green in color, light in greyscale) at the considered point.
  The two black dots indicate where the trajectories begin.
  Note that side (a) shows a few precessions of $\avg{m_x}$,
  while side (b) shows only one precession.
  This is just to avoid the overlapping of the line.
}
\end{figure}
Fig. \ref{fig_enpump}(a) provides an example of this phenomenon.
The plot shows the initial part of the trajectory of the spatially
averaged normalized magnetization
$\langle \vv{m} \rangle = \langle \vv{M}/\Msat \rangle$ in
the plane $yz$ for the case $\pinang = 20\dg$ of Fig.~\ref{fig_jdyn}.
The points of the trajectory where the energy of the system is increasing,
$\derivp{E}{t} > 0$, are coloured in red; the points where it is decreasing,
$\derivp{E}{t} < 0$, are coloured in green.
The line thickness is proportional to $|\derivp{E}{t}|$: a thick red line
corresponds to maximum energy increase, while a thick green line to maximum
energy decrease.
The plot shows that once the current is applied, the energy starts increasing
while the magnetization moves in the negative $y$ direction.
At this point, however, the pinning is strong enough to oppose the
current-driven dynamics and pulls the magnetization back, giving rise to a
motion spiralling towards a new equilibrium configuration.
During the precession, the STT pumps energy in and out, in an alternate
fashion. Overall, the current produces changes in $\vv{m}$ of the order
of $4 \times 10^{-4}$ radians.

A completely different situation is shown in Fig. \ref{fig_enpump}(b),
which shows again the dynamics of $\langle \vv{m} \rangle$, but
for a DW moving through the rotationally symmetric barrier
(case $\theta_{\mathrm{P}} = 0\dg$ in Fig. \ref{fig_jdyn}).
The plot shows that the energy is always pumped in (and never pumped out).
Moreover, the rate with which this happens (line thickness) increases in time.
This is due to the compression of the DW, which causes a stronger
STT effect (higher values of $\derivp{\vvm}{x}$ in
Eq. \eqref{eq:llgzl_explain}).

The analysis we have carried out demonstrates that it is of fundamental
importance to find systems and situations which allow the STT to coherently
pump energy into the system.
The effectiveness of the STT in the case of rotationally symmetric
and asymmetric barriers depends on whether or not the STT points along
the optimal direction for energy pumping (i.e. opposite to the damping).

\section{Energy barrier}
It may be useful at this point to give a more quantitative indication
on the height of the barriers which can be penetrated thanks to the mechanism
described in this paper.
In particular, we would like to find out, given a rotationally symmetric
barrier with a given height $\Delta E$, what the required critical current
density is in order for the DW to pass through it.

\begin{figure}
\includegraphics[width=8.0cm]{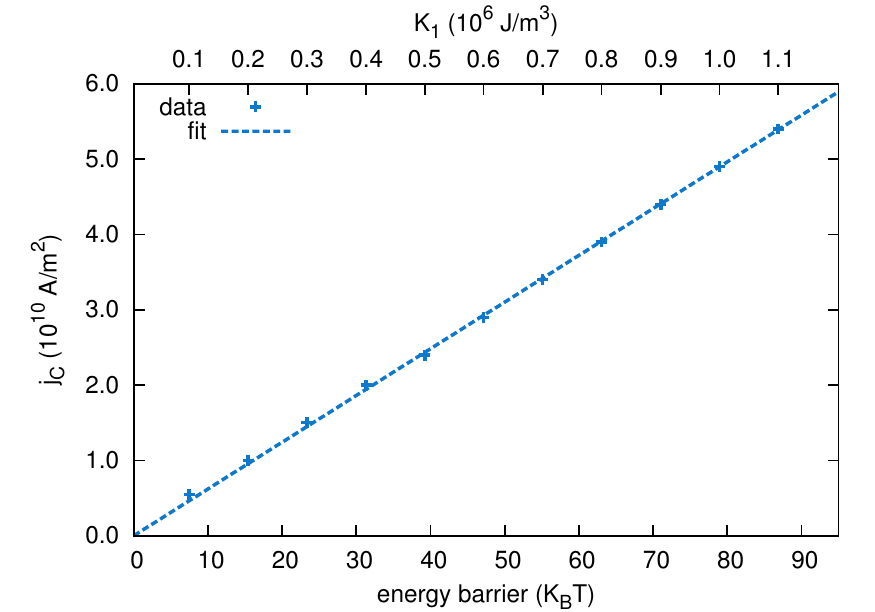}
\caption{\label{fig_jcrit_vs_k1} (Color online)
  Study of the critical current, $\jcrit$ required to push the DW through
  a rotationally symmetric barrier, as a function of the barrier height
  ($\Delta E$).
  The top $x$ axis shows the values of the anisotropy constant $K_1$
  corresponding to the values of $\Delta E$ shown on the bottom axis.}
\end{figure}

We perform simulations similar to the ones described previously,
but here we fix $\pinang=0\dg$ and change the value for the
pinning anisotropy $K_1$ from $0.1\times10^6\,\mathrm{J/m^3}$ to
$1.1\times10^6\,\mathrm{J/m^3}$ in steps of $0.1\times10^6\,\mathrm{J/m^3}$.
For every simulation we also calculate the energy height $\Delta E$
as the difference $\Delta E = E_{\mathrm{max}} - E(0)$, where $E(0)$
is the energy of the system at the beginning of the simulation, while
$E_{\mathrm{max}}$ is the maximum value reached throughout the simulation.
The values of the critical current, $\jcrit$, are plotted in
Fig.~\ref{fig_jcrit_vs_k1} as function of the barrier height, $\Delta E$.
The corresponding values for $K_1$ are also shown at the top of the plot.
We note that there is a linear relationship between all the three
quantities involved: $\jcrit$, $\Delta E$ and $K_1$.
In particular, $\Delta E$ and $K_1$ (bottom and top $x$ axis)
are linearly proportional and follow the relation
$\Delta E (K_1) = V \, K_1$ where $V$ is a parameter with dimensions
of a volume and $V = (326.4 \pm 0.4) \, \mathrm{nm^3}$ (from a least squares fit).
This means that --- as one may expect --- a doubled value for $K_1$
corresponds to twice-as-high energy barrier, $\Delta E$.

The linear relation between $\jcrit$ and $\Delta E$ is somewhat more difficult
to understand intuitively.
Fig.~\ref{fig_jcrit_vs_k1} shows, however, that the data can be well
fitted (see dashed curve) against the function,
\begin{equation}
\jcrit(\Delta E) = \tilde{B} \, \Delta E.
\label{eq:jcrit_of_de}
\end{equation}
In particular, we get
$\tilde{B} = (1.498 \pm 0.006) \times 10^{29} \, \mathrm{A/(m^2 J)}$.
Or --- in terms of thermal energy --- it is necessary to apply a fully
polarized current density $j = P\japp$ of $2.48\times10^{10}\,\mathrm{A/m^2}$
in order to push the DW through a potential barrier with height $40\,\kbt$.
Assuming a spin polarization $P=0.4$,
we find $\japp = 6.20\times10^{10}\,\mathrm{A/m^2}$.

We now multiply both sides of equation \eqref{eq:jcrit_of_de} by the cross
sectional area of the nanowire, $\pi R^2$:
\begin{equation}
\Icrit(\Delta E) = B \, \Delta E,
\label{eq:Icrit_of_de}
\end{equation}
which relates the critical current $\Icrit = \pi R^2 \, \jcrit$ to $\Delta E$.
We have
$B = \pi R^2 \, \tilde{B} = (1.177 \pm 0.005) \times 10^{13}\,\mathrm{A/J}$.

\section{Analytical considerations}
In this section we derive a simple analytical model with the aim of
understanding the linear proportionality between the current density, $\jcrit$,
required to push a DW through a potential barrier,
and the barrier height, $\Delta E$ (as seen in Fig.~\ref{fig_jcrit_vs_k1}).
We start from the analytical model developed in
Ref.~\onlinecite{Franchin2008prb}
(the analytical derivation below is heavily based on this previous work).
In particular, we consider the situation where the pinning potential
is infinitely high and the DW does not penetrate into the barrier,
but rather compresses against it, thus accumulating a certain amount
of exchange/compression energy.
It is reasonable to expect that if the applied current can accumulate an
amount of ``compression energy'', $\Delta E$, in the case of an infinite barrier,
then it may also be able to push the DW through a finite energy barrier
of the same height.
We stress that in this context we are just trying to get to a qualitative
understanding of the phenomenon, rather than a quantitative analytical model.

First, we write down the energy of the DW, using the one dimensional model,
the coordinate system and nomenclature of Ref.~\onlinecite{Franchin2008prb},
$$
U =
A \int_0^L \left[ (\derivp{\theta}{x})^2
             + \sin^2 \theta \, (\derivp{\phi}{x})^2 \right]
       \mathrm{d} x.
$$
$U$ has the units of energy per cross-sectional area.
$A$ is the exchange coupling constant, $L$ the DW width, $\theta$ and $\phi$
are spherical coordinates for the magnetization.
In this expression, we consider only the exchange energy and neglect
the contribution of the magnetostatic field.
The term $\derivp{\phi}{x}$ is usually smaller\cite{Franchin2008prb}
than $\derivp{\theta}{x}$, and we neglect it:
$$
U = \frac{A}{L} \int_0^1 (\derivp{\theta}{u})^2 \, \mathrm{d} u
  = -\frac{A}{L} \int_0^{\pi} \derivp{\theta}{u} \, \mathrm{d} \theta,
$$
where we also changed the variable of integration twice,
from $x$ to $u = x/L$ and from $u$ to $\theta$.
In the high current regime\cite{Franchin2008prb},
$\derivp{\theta}{u} = -\frac{3V}{\alpha} \, \sin \frac{\theta}{2}$,
and,
$$
U = \frac{3 A V}{L \alpha} \int_0^{\pi} \sin \frac{\theta}{2} \, \mathrm{d} \theta
  = \frac{6 A V}{L \alpha}.
$$
Since, $V = \frac{L}{\gamma C}\,v$, we have:
\begin{equation}
U = \frac{6 A V}{L \alpha} = \frac{3 \mu_0 \Msat}{\gamma \alpha} \, v
  = \frac{3 \mu_0 \mu_B}{e} \, \frac{j}{\gamma \alpha}
\label{eq:jc_vs_en_slope}
\end{equation}
For the low current regime the DW deformation is negligible and we get
the energy of a relaxed transverse wall of length $L$, $U_0 = A \pi^2/L$.
In summary, the present analysis is valid only in the high current regime,
i.e. for currents
$$
j = P \japp \gg j_0 = \frac{2e \gamma}{\mu_0 \mu_B} \, \frac{\alpha A}{L}.
$$
In this regime, the DW compresses accumulating
an exchange energy $U$ per unit of cross sectional area,
$$
U = \frac{6}{\pi^2} U_0  \, \frac{j}{j_0}, \quad \mathrm{for}\,j \gg j_0.
$$
The energy pumped in by the current is $\Delta U = U - U_0$,
\begin{equation}
\Delta U = U_0 \left( \frac{6}{\pi^2} \, \frac{j}{j_0} - 1 \right),
  \quad \mathrm{for}\,j \gg j_0.
\label{eq:de_from_j}
\end{equation}
In our system we have $j_0 = 0.026 \times 10^{12}\,\mathrm{A}/\mathrm{m}^2$
(we have assumed a DW length $L = 30\,\mathrm{nm}$).
We expect Eq. \eqref{eq:de_from_j} to hold only when the current
exceeds this value.
When $j \sim j_0$, on the other hand, we may still have important
STT effects, but getting to an analytical expression would require
to solve exactly the system of equations (14) in Ref.~\onlinecite{Franchin2008prb}.
We can now compare the value obtained for $B$ (see Eq. \eqref{eq:Icrit_of_de})
from the fit, $B = (1.177 \pm 0.005) \times 10^{13}\,\mathrm{A/J}$, with
$B = 1.013 \times 10^{13}\,\mathrm{A/J}$, as obtained from the equation below
(derived from Eq. \eqref{eq:jc_vs_en_slope}),
$$
B = \frac{e \gamma \alpha}{3 \mu_0 \mu_B}.
$$
Note that $B$ does not depend on the considered geometry,
meaning that the total depinning current, $\Icrit$, does not depend on the size
of the nanowire. In larger nanowires then lower current densities can be used
to overcome barriers of the same height in energy.
For example, the current density, $\japp = 6.20\times10^{10}\,\mathrm{A/m^2}$,
required to push a transverse domain wall through a barrier of $40\,\kbt$
in a nanowire with diameter $10\,\mathrm{nm}$, should be reduced by a factor
$9$ in nanowires with diameter $30\,\mathrm{nm}$.
The optimization of the system geometry necessary to achieve
lower current densities and lower switching times are left to future
investigations.

In summary, Eq. \eqref{eq:de_from_j} shows that the amount of exchange energy
the DW accumulates by compressing against the infinite potential barrier
is proportional to the applied current density.
Fig.~\ref{fig_jcrit_vs_k1} is then easily understood, assuming the linear
behaviour is maintained in the case of finite barrier.

\section{Summary}
We have studied a transverse DW in a cylindrical nanowire and
its motion through a barrier,
modeled as a pinning potential on the magnetization.
We determined the critical fields and current densities required
to push the DW through the barrier for various directions
of the pinning potential.
We found that the critical applied field decreases as the pinning direction
gets orthogonal to the nanowire axis.
On the contrary, the critical current density increases by more than a
factor 130 when the pinning direction gets orthogonal to the nanowire axis,
meaning that the DW can penetrate the barrier much more easily when
the pinning potential is aligned along the axis of the wire,
rather than being orthogonal to it.

This study gives further insights into the extraordinary properties
of transverse DWs in cylindrical nanowires and motivates experimental
investigations on these systems.

\begin{acknowledgments}
We thank Guido Meier for valuable discussions and for sharing
experimental data prior to publication.
The research leading to these results has received funding from the
European Community's Seventh Framework Programme (FP7/2007-2013) under
Grant Agreement n$^{\circ}$ 233552,
and from EPSRC (EP/E040063/1 and EP/G03690X/1).
\end{acknowledgments}

\end{document}